\documentclass{vgtc}                          

\ifpdf
  \pdfoutput=1\relax                   
  \usepackage{graphicx}                
  \DeclareGraphicsExtensions{.pdf,.png,.jpg,.jpeg} 
\else
  \usepackage{graphicx}                
  \DeclareGraphicsExtensions{.eps}     
\fi%

\graphicspath{{figures/}{pictures/}{images/}{./}} 

\usepackage{microtype}                 
\PassOptionsToPackage{warn}{textcomp}  
\usepackage{textcomp}                  
\usepackage{mathptmx}                  
\usepackage{times}                     
\usepackage{cite}                      
\usepackage{tabu}                      
\usepackage{booktabs}                  

\onlineid{0}

\vgtccategory{Research}

\vgtcinsertpkg

\title{Exploring Effectiveness of Explanations for Appropriate Trust: Lessons from Cognitive Psychology
}

\author{Ruben S. Verhagen*\thanks{e-mail: r.s.verhagen@tudelft.nl; *Author sharing first authorship.} %
\and Siddharth Mehrotra*\thanks{e-mail: s.mehrotra@tudelft.nl; *Author sharing first authorship.} %
\and Mark A. Neerincx\thanks{e-mail:m.a.neerincx@tudelft.nl; Secondoary affiliation: TNO, Soesterberg, The Netherlands}
\and Catholijn M. Jonker\thanks{e-mail:c.m.jonker@tudelft.nl; Secondoary affiliation: Leiden University, The Netherlands}
\and Myrthe L. Tielman\thanks{e-mail:m.l.tielman@tudelft.nl}}
\affiliation{Delft University of Technology, The Netherlands}

\abstract{The rapid development of Artificial Intelligence (AI) requires developers and designers of AI systems to focus on the collaboration between humans and machines. AI explanations of system behavior and reasoning are vital for effective collaboration by fostering appropriate trust, ensuring understanding, and addressing issues of fairness and bias. However, various contextual and subjective factors can influence an AI system explanation's effectiveness. This work draws inspiration from findings in cognitive psychology to understand how effective explanations can be designed. We identify four components to which explanation designers can pay special attention: perception, semantics, intent, and user \& context. We illustrate the use of these four explanation components with an example of estimating food calories by combining text with visuals, probabilities with exemplars, and intent communication with both user and context in mind. We propose that the significant challenge for effective AI explanations is an additional step between explanation generation using algorithms not producing interpretable explanations and explanation communication. We believe this extra step will benefit from carefully considering the four explanation components outlined in our work, which can positively affect the explanation's effectiveness.} 

\CCScatlist{
  \CCScatTwelve{Human-centered computing}{Visu\-al\-iza\-tion}{Visu\-al\-iza\-tion techniques}{};
  \CCScatTwelve{Human-centered computing}{Visu\-al\-iza\-tion}{Visualization design and evaluation methods}{}
}

\begin{document}



\maketitle

 \section{Introduction} 
Humans and Artificial Intelligence (AI) systems are increasingly collaborating on tasks ranging from the medical to the financial domain. For such human-machine collaboration to be effective, mutual understanding and trust are of paramount importance \cite{johnson2019no, klien2004ten, salas2005there}. AI explanations are a crucial and powerful way to increase human understanding and trust in the system. Explanations can help in explaining the decisions and inner workings of the "black-box" data-driven machine learning algorithms and actions \& reasoning of goal-driven agents \cite{anjomshoae2019explainable, gunning2017explainable, pecox, guidotti, langley2017explainable, miller2019explanation}. 

Unfortunately, AI systems often lack transparency and explainability, providing a broad call for explainable AI (XAI). For example, EU GDPR requires organizations that deploy AI systems to provide relevant information to affected people about the inner workings of the algorithms \cite{voigt2017eu}. Furthermore, in addition to helping people understand AI systems, AI explanations can also contribute to identifying and addressing issues of fairness and bias, which are difficult to do without. Explanations can also help form appropriate trust in the AI system. For example, Dodge et al. \cite{dodge2019explaining} pointed out that when people trust the explanation, they are more likely to trust the underlying ML system.

A common challenge in popular explanation methods such as LIME (Local Interpretable Model-agnostic Explanations) by Ribeiro et al. \cite{lime}, SHAP (Shapley Additive Explanations) by Lundberg \& Lee \cite{lundberg}, and global-local explanation by Lundberg et al. \cite{lundberg2020local} is the effectiveness of the explanation to help users calibrate their trust and increase understanding of the explanation. Various aspects can impact the effectiveness of AI explanations, such as the user's domain, system expertise or cognitive and perceptual biases. However, this list is far from conclusive. Therefore it is crucial to investigate the following two research questions:
\begin{enumerate}
    \item \textit{What should be the content of the explanation to make it effective?}
    \item \textit{How can the visual delivery or design of the explanation make it effective?} 
\end{enumerate}
\section{Explanation Effectiveness}
In this work, we take inspiration from cognitive psychology, design, and data visualization techniques to explore ways how effective explanations can be designed. A large amount of work in cognitive psychology focuses on explanations in human-human interactions, what makes them effective and how context can affect the same. For example, Lombrozo \cite{lombrozo2006structure} provides two properties of the structure of explanations that helps in reasoning, (1) explanations accommodate novel information in the context of prior beliefs, and (2) do so in a way that fosters generalization. The author showcase that explanations provides a unique window onto the mechanisms of learning and inference in human reasoning.

David B. Leake in his book ``Evaluating Explanations" from cognitive psychology theories has described how context involving both explainer beliefs and goals can help in deciding an explanation's goodness \cite{leake}. Similarly, Khemlani et al. shows how mental models represent causal assertions, and how these models underlie deductive, inductive, and abductive reasoning yielding effective explanations \cite{khemlani2014causal}. Since explanations determine how humans understand the world in fundamental terms, Tworek and Cimpian helps in exploring human biases towards the inheritance of judgments in people’s explanations over socio-moral understanding \cite{tworek}.

The previously mentioned works and recent research in human-AI interaction \cite{setlur2022functional,nauta2022anecdotal,zhou2021evaluating,vilone2021notions,hearst2019toward} helps us in exploring our two research questions based on four components (Perception, Semantics, Intent, and User \& Context). These components are uniquely visible at the intersection of studies in human-AI interaction and cognitive psychology. We will now describe these components in detail.

\subsection{Perception}
Perception in XAI can refer to the set of mental processes we use to make sense of a given explanation by an AI system. Perception is also frequently related to the modality in which explanation is presented and how our brain interprets it. For instance, example-based explanations are useful in cases where it's hard to explain AI reasoning. Here, developers can create an explanation in the form of a real-life example or a story to depict the system's processing, in line with how we as humans are primed to relate to real-life examples or episodes. Empirical research has demonstrated that when explanations take the form of a story, they can help humans in decision making \cite{jonassen2012designing}. According to Rumelhart \cite{rumelhart1979understanding}, stories often contain initiating events, goals, actions, consequences, and accompanying states in a particular causal configuration. Knowing the structure of stories can allow humans to form appropriate trust concerning the perception of the explanation.

For saliency maps, categorical model confidence visualizations, n-best visualizations, and related ways of visual explanations, we propose that it is important to consider how the human brain is organized to see structures, logic, and patterns. Our brain organization helps us as humans to make sense of the world. A seminal work by Wertheimer \cite{wertheimer1938gestalt} introduced seven Gestalt principles of visual perception in the form of heuristics. We now briefly describe those principles in relation to visual explanations in XAI:
\begin{enumerate}
    \item \textbf{\textit{figure-ground}}: visualization of the explanation can display objects as being either in the foreground or the background. For example, an identified object with a higher confidence score can be displayed in the foreground, leaving every other object in the background.
    \item \textbf{\textit{similarity}}: clustering of similar objects can be shown to the user such that the user can easily group similar objects together. For example, we can use a variety of design elements, like color and organization, to establish similar groups. 
    \item \textbf{\textit{proximity}}: for pixel-wise importance in saliency map of human face detection, calculated using integrated gradients; explanations can focus on (a) the areas around the eyes, (b) lines anywhere on the face, and (c) regions around the mouth and nose. This relative nearness of the objects can strongly influence the user's understanding of the explanation.
    \item\textbf{\textit{common region}}: the principle of common region states that when objects are located within the same closed region, humans perceive them as being grouped together. Therefore, for any two distinct objects, a visual explanation can put them in different regions and vice-versa. 
    \item \textbf{\textit{continuity and closure}}: continuity refers to the arrangement of elements while closure refers to identification of recognizable patterns. We can apply these principles in explanations presented in a commonly used dashboard. For example, these principles explain why only two axes, rather than a full enclosure, are required on a graph to define the space in which the data appears. Also, it can become obvious which groups belong to the subgroup when the hierarchy is expanded.
    \item \textbf{\textit{focal point}}: the premise of the focal point principle is that anything visually outstanding first captures and retains the viewer's attention. In terms of explanations, assigning each point a score reflecting its contribution to the ML model-recognition loss can follow this principle. The saliency map can then explicitly explain which points are the key to model recognition, as explained by Zheng et al. \cite{zheng2019pointcloud}.
\end{enumerate}
Zena O’Connor's contemporary work on colour, and contrast in combination with gestalt theories provide the effectiveness of visual communication design in graphical design, which can be helpful for the XAI community \cite{o2015colour}. For example, the author examined the role of colour and contrast within the context of Gestalt theories of perception. The author also showcased how colour and contrast not only help to distinguish contours, detail, and depth, but they also help to attract and divert attention, thereby clearly delineating key areas of text. This result can help design explanations when multiple models are used to describe a decision by an AI system.
\subsection{Semantics}
Semantics in XAI refers to how AI system designers can convey the meaning in language to their users. This meaning of the language can help users get a grip on what the expressions of a natural language contribute to the overall process of interpretation. Cann et al. \cite{Cann2009-CANSAI-2} provide three basic types of sense relations to clearly communicate the meaning of the language: 
\begin{enumerate}
    \item \textit{Synonymy}: sameness of sense; for example, X's pullover is yellow v/s X's sweater is yellow.
    \item \textit{Hyponymy}: sense inclusion; for example, X wants a hamburger v/s X wants to eat a hamburger.
    \item \textit{Antonymy}: oppositeness in sense; for example, This water is cold v/s This water is not hot.
\end{enumerate}
It becomes essential to be aware of these sense relations when designing explanations for AI systems because as soon as AI system designers start probing what concept of meaning they should articulate, they may slip away into a quag of open-endedness. 

Researchers have focused on resolving linguistic ambiguities in visual data communication. For example, Setlur et al. \cite{setlur2019inferencing} provide heuristics to resolve partial utterances based on syntactic and semantic constraints. Similarly, Law et al. \cite{law2021causal} have proposed ways such as encouraging skepticism and open-mindedness to reduce the illusion of causality when using question-answering systems. Similarly, Gaba et al. \cite{gaba2022comparison} have provided guidelines for designing natural language interfaces and recommendation tools to better support natural language comparisons in visual communication. These guidelines include use of basic charts as reasonable responses for a variety of comparisons, providing necessary information useful for the comparisons, and exposing the provenance of how implicit entities are such as “cheap” and “best-selling”.

Humans often adapt cognitive approaches to explanations where any ambiguity and vagueness can affect human-AI communication. This fuzziness in communication can result in distrust in the AI system and can be disastrous in high-risk scenarios where lives can be at stake. Taken into consideration the fuzziness of explanations, Nauta et al. \cite{nauta2022anecdotal} provide a summary of Co-12 properties on explanation quality. The properties related to semantics (and perception) fall in the category of content focusing on correctness, completeness, contrastivity, continuity, consistency, and covariate complexity. These properties provide an aggregated view of what to evaluate in an explanation. We believe that an explanation's cohesiveness ties all these properties together. In cohesiveness, one can focus on creating semantically resonant explanations that enrich human understanding. When an explanation is cohesive it can also help in preserving users’ mental models and memory. For example, providing infographic information with confidence scores based on reasoning with facts from verified and trusted sources can help make an explanation cohesive. Furthermore, exploring concepts around scaling or hiding information can create the right balance for explanation effectiveness.

Apart from focusing on effective explanations qualities, it is also essential to consider how data literacy can impact the perceived effectiveness of an explanation. Wolf's and Gottwald's popular work on \textit{Tales of Literacy} sheds light on how digital literacy is differently represented in the human brain \cite{wolf2016tales}. The authors present "vocabulary" of language and reading development to appreciate better the critical importance of specific literacy development aspects with digital systems worldwide. On a similar note, Setlur and Cogley \cite{setlur2022functional} provide an overview of data literacy and how it forces us to think about how everyone navigates the world. Authors call out to researchers working in this domain to focus on cultural differences in data literacy and how people around the globe can interpret an explanation in multiple ways.

Any recommendation in form of an explanation by an AI system is based on a probabilistic model. One of the challenges for effective explanations is identifying a baseline for communicating these probabilities. Based on the work by Wolf \cite{wolf2008proust}, we propose that ability of neural networks in the brain helps and hinders humans in their attempt to read and process information related to probabilities. Since our way of processing information differs at the individual level, so does our understanding of an explanation. Therefore, it becomes vital to consider diverse populations and the moral components of the design and use of visualizations for explanations in an easy to understand manner \cite{correll2019ethical}.

We believe that AI systems will not be explainable without first addressing the scarcity of semantics in explanations. Browne and Swift  \cite{browne2020semantics} highlights that \textit{``producing satisfactory explanations for deep learning systems will require that we find ways to interpret the semantics of hidden layer representations in deep neural networks"}. There have been a few attempts looking at semantics in explanations, such as by Jin et al. \cite{jin2019towards} exploring hierarchical visualization of compositional semantics to help users create trust in deep neural networks. Similarly, Jacob et al. \cite{jacob2021steex} investigating how users can guide the generation of counterfactual explanations by specifying a set of semantic regions of the query image. Although both reported works show promising results, extracting semantic explanations from hidden units is far from a solved problem.

\subsection{Intent}
\begin{figure*}[h]
\includegraphics[width=1.0\textwidth]{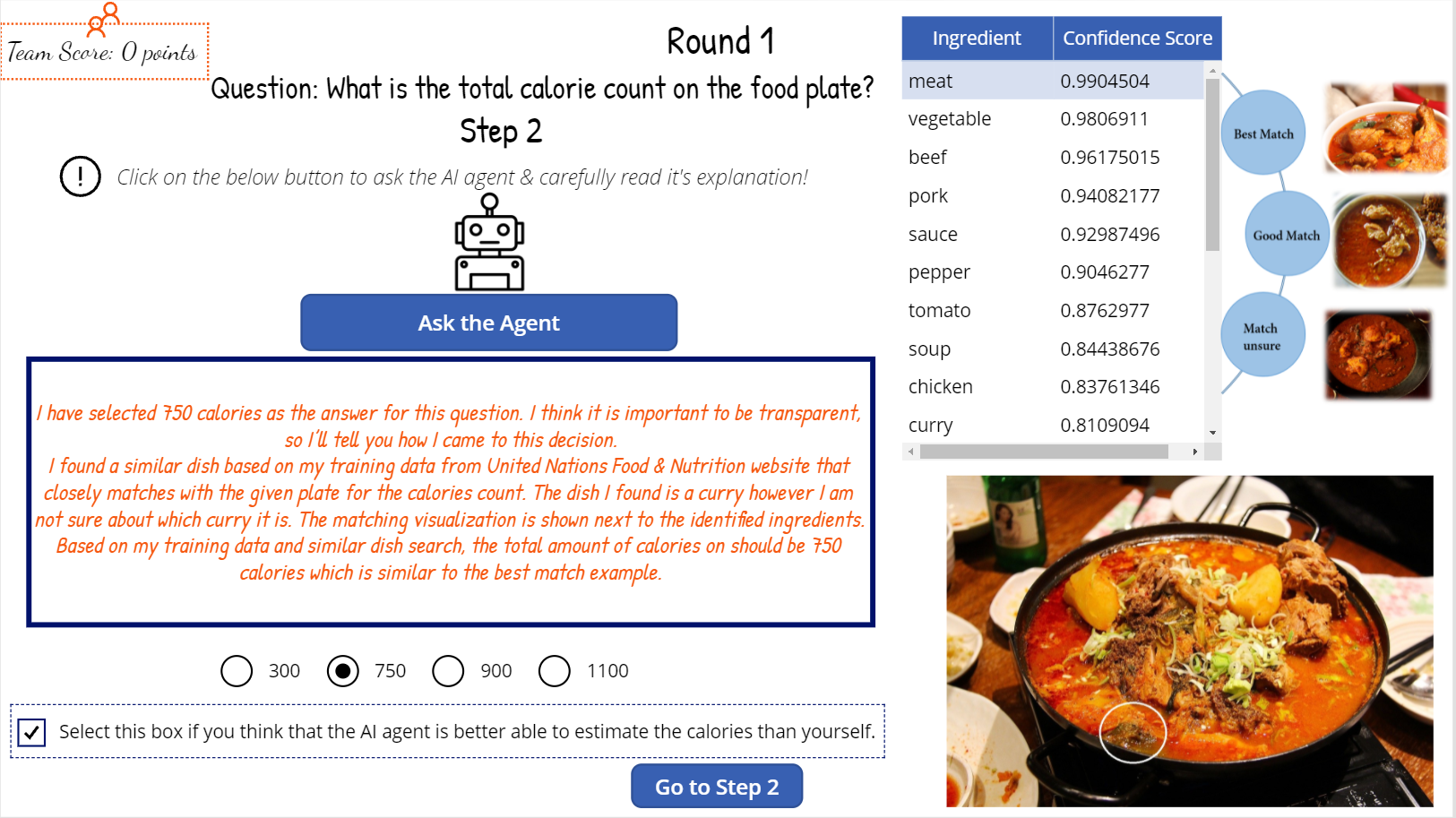}
    \caption{In this snapshot of a user study, an AI agent provides an estimation of calories of the dish (shown in bottom right) out of four available options (shown in bottom left). Here the estimation of calories is accompanied by confidence scores of each identified ingredient(s) in a form of a table. Also, the AI agent provides both textual and visual explanations for its reasoning of selecting an option for calories estimation.}
  \label{fig:teaser}
 \end{figure*}
The intent in XAI refers to the purpose of the explanation, ideally balancing semantics with goals. Therefore, intent can heavily influence the perception and semantics components when designing explanations. For example, explanations aimed at convincing users to engage in a healthier lifestyle should take different forms than explanations aimed at improving firefighters' situation in high stake decisions. For recommending food options to a user to keep a healthy lifestyle, explanations can be designed to (a) convince users with confidence estimates of the nutrition of food, (b) predict the outcome of eating the healthy food, or (c) showcase low confidence when a system can't recommend a healthy diet based on available food option to foster appropriate trust. In contrast, explanations aimed at improving the situation and moral awareness of firefighters can explain the potential consequences of the most important situational features to consider when making a decision. 

From the literature of cognitive psychology, JS Bruner \cite{bruner1974communication} sheds light on Macfarlane's work, where authors studied how mothers are irresistibly imputing intent to the cries, gestures, expressions, and postures of newborns. As communication presupposes intent in communicating an objective, one can learn so much just by understanding a mother's intent to understand her child. Bruner also highlights that in interpreting the infant's communicative intent - correctly or incorrectly - the mother has a rich variety of cues to use, and so too for the child. Relating this to the XAI context, we propose that the AI system designers need to adapt their system designs to match the user's intent based on available cues, similar to how a mother does for her child, as outlined by Bruner.

Neerincx et al. and Anjomshoae et al. \cite{pecox, anjomshoae2019explainable} provide three explanation phases to convey intent of an explanation: generation, communication, and reception. Explanation \textit{generation} mostly concerns with the explainability methods used to construct relevant elements and information to share, such as SHAP or LIME explainers. Explanation \textit{communication} refers to the form and content and \textit{reception} relates to how the receiving user utilizes and understands the explanation. In the work by Miller \cite{miller2019explanation}, the author did emphasize the crucial role of intention in explanations, arguing that determining the goal of an explanation is key to providing a good explanation. We resonate with this statement and propose that the current field of XAI can pay more attention to explanation intent. Specifically, intent can guide explanation communication (i.e., content\slash semantics and form\slash perception), which can show its benefits in explanation reception. 
\subsection{User and Context}
We believe that defining explanation intent includes considering both user and context. Recent works emphasize the importance of personalized and context-aware explanations \cite{anjomshoae2019explainable, arrieta2020explainable, miller2019explanation}, as it is believed that they improve explanation effectiveness. For example, the same explanation used to convince user A to live healthier will likely not work if user B has a completely different value profile or personality. Likewise, when a firefighter is under extreme time pressure, visual explanations are probably much more effective than verbal ones. Taken together, we believe that simply defining explanation intent as to inform, convince, or support will not suffice. Instead, explanation designers can define the goal by combining explanation intent, receiver\slash user, and context (e.g., \textit{convince conservative user X to start exercising more}). 


\section{Use case example}
Wang and Yin provided three desiderata for designing effective AI explanations \cite{wang2022effects}. These desiderata include (a) designing explanations to improve people’s understanding of the AI model, (b) helping people recognize the uncertainty underlying an AI prediction, and (c) empowering people to trust the AI appropriately. In the following example of an AI explanation, we follow these desiderata together with the components of explanations that we discussed earlier.

Our use case example showcases a snapshot of the “estimating calories on a plate” user study, see Figure \ref{fig:teaser}. We designed the task around nutrition as an approachable domain for everyone. In this user study, an AI agent helps the user to estimate the food calories with the help of explanations. The AI agent estimates the calories of a dish shown in the right corner of the figure. The agent also explains its confidence scores (based on an AI-driven food recognition model). In addition, the categorical visualization for each ingredient is displayed as a form of visual explanation. The user must decide among the four options to select the closest calorie count based on the AI agent's suggestion and get a +10 score for a correct selection, whereas a wrong selection costs -10 points.

In the above use case, the confidence scores of ingredients are generated using a machine learning classifier in real-time. The visual explanations are based on categorical visualization (Figure \ref{fig:teaser}, right) inspired by the example provided in the Google PAIR guidebook \cite{google}. These visual explanations were designed keeping in mind the four components discussed in this paper as follows: 
\begin{enumerate}
    \item In our use case, \textbf{perception} refers to different categorizations such as best, good, or unsure match by providing clear meaning of the categorizations to the user.
    \item For \textbf{intent}, we decided to use a single image for each categorization to convey the actual semantics we perceived as system designers. The text-based explanations in the figure are handcrafted following the notion of situation vignettes \cite{strack2011personal}. Here every explanation starts with the intent the AI agent wants to convey, followed by the perception of the authority (in this case United Nations Food \& Nutrition website) on which the suggestion is based. 
    \item The next part of the explanation is to clearly deliver the \textbf{semantics} of the reasoning behind the suggestion to the user in non-ambiguous language. For semantics, we adopted Rumelhart's \cite{rumelhart1979understanding} guidelines and followed gestalt heuristics for designing visual explanation. 
    \item The \textbf{user and context} are also taken into account by relating the explanation with the closest dish in the database and providing logic-based reasoning that users can adhere to. Furthermore, the combination of visual and textual explanation is designed to elicit an appropriate amount of trust that the AI agent can foster, distinct from promoting the trust in the agent.
\end{enumerate}
Our method of showcasing explanations is a way to incorporate the components that we have discussed in this work. This approach has known limitations, such as scaling the explanations with complex data or crafting these explanations in real-time. We believe that these limitations can be overcome with further advancements in the field of XAI with methods such as ensemble ML, feature-based, rule-based, and training-data based explanations.
\section{Conclusion}
Previous work has shown that the efficiency of human-AI interaction can be improved when AI systems explain their behavior and reasoning. However, how to create explanations that are both intuitive and easily understandable by users is a challenging task. In this work, we have discussed four components of designing effective explanations based on insights from cognitive psychology. Furthermore, we provided an example to incorporate those components in creating a textual and visual explanation for a use-case example. With this work, we propose the need to provide emphasis on these components to make explanations effective. We also showed how communicating results in an interpretable manner needs to be as intuitive as dialogue. However, we must note that explanations can increase the chance that humans will accept the AI’s recommendation, regardless of its correctness. Therefore, explanations should strive to create appropriate trust rather than helping promote trust in the AI system. Also, we highlight that it is important to avoid explanations biasing users towards certain actions or decisions, especially in high-stakes decision-making tasks.  
\acknowledgments{
The authors wish to thank the three anonymous reviewers for their constructive comments and review of the manuscript.}
\bibliographystyle{abbrv-doi}

\bibliography{template}
\end{document}